\documentclass[12pt,preprint]{aastex}

\begin{document}

\title{Testing Models of Magnetic Field Evolution of Neutron Stars with the Statistical Properties of Their Spin Evolutions}

\author{Shuang-Nan Zhang\altaffilmark{1,2}, Yi Xie\altaffilmark{1}}
\altaffiltext{1}{National Astronomical Observatories, Chinese Academy of Sciences, Beijing, 100012, China;
zhangsn@ihep.ac.cn}\altaffiltext{2}{Key Laboratory of Particle Astrophysics, Institute of High Energy Physics,
Chinese Academy of Sciences, Beijing 100049, China}

\begin{abstract}
We test models for the evolution of neutron star (NS) magnetic fields (B). Our model for the evolution of the NS
spin is taken from an analysis of pulsar timing noise presented by Hobbs et al. (2010). We first test the
standard model of a pulsar's magnetosphere in which B does not change with time and magnetic dipole radiation is
assumed to dominate the pulsar's spin-down. We find this model fails to predict both the magnitudes and signs of
the second derivatives of the spin frequencies ($\ddot{\nu}$). We then construct a phenomenological model of the
evolution of $B$, which contains a long term decay (LTD) modulated by short term oscillations (STO); a pulsar's
spin is thus modified by its B-evolution. We find that an exponential LTD is not favored by the observed
statistical properties of $\ddot{\nu}$ for young pulsars and fails to explain the fact that $\ddot{\nu}$ is
negative for roughly half of the old pulsars. A simple power-law LTD can explain all the observed statistical
properties of $\ddot{\nu}$. Finally we discuss some physical implications of our results to models of the
$B$-decay of NSs and suggest reliable determination of the true ages of many young NSs is needed, in order to
constrain further the physical mechanisms of their $B$-decay. Our model can be further tested with the measured
evolutions of $\dot{\nu}$ and $\ddot{\nu}$ for an individual pulsar; the decay index, oscillation amplitude and
period can also be determined this way for the pulsar.

\end{abstract}


\section{Introduction}           
Many studies on the possible magnetic field decay of neutron stars (NSs) have been done previously, based on the
observed statistics of their periods ($P$) and period derivatives ($\dot{P}$). Some of these studies relied on a
population synthesis of pulsars (Holt \& Ramaty 1970; Bhattacharya et al. 1992; Han 1997; Regimbau \& de Freitas
Pacheco 2001; Tauris \& Konar 2001; Gonthier et al. 2002; Guseinov et al. 2004; Aguilera et al. 2008; Popov et
al. 2010); however, no firm conclusion can be drawn on if and how the magnetic fields of NSs decay (e.g.,
Harding \& Lai 2006; Ridley \& Lorimer 2010; Lorimer 2011). Alternatively some other studies used the spin-down
or characteristics ages, $\tau_{\rm c}=(P-P_{0})/2\dot{P}$ ($P_{0}$ is the initial spin period of the given
pulsar), or $\tau_{\rm c}=P/2\dot{P}$ if $P\gg P_{0}$, as indicators of the true ages of NSs and found evidence
for their dipole magnetic field decay (Pacini 1969; Ostriker \& Gunn 1969; Gunn \& Ostriker 1970). However, as
we have shown recently (Zhang \& Xie 2011), their spin-down ages are normally significantly larger than the ages
of the supernova remnants physically associated with them, which in principle should be the unbiased age
indicators of these NSs (Lyne 1975; Geppert et al., 1999; Ruderman 2005). It has been shown that this age
mismatch can be understood if the magnetic fields of these NSs decay significantly over their life times, since
the magnetic field decay alters the spin-down rate of a NS significantly, and thus cause the observed age
mismatch (Lyne 1975; Geppert et al. 1999; Ruderman 2005; Zhang \& Xie 2011).

Pulsars are generally very stable natural clocks with observed steady pulses. However significant timing
irregularities, i.e., unpredicted times of arrival of pulses, exist for most pulsars studied so far (see Hobbs
et al. 2010, for an extensive reviews of many previous studies on timing irregularities of pulsars). The timing
irregularities of the first type are ``glitches", i.e., sudden increases in spin rate followed by a period of
relaxation; it has been found that the timing irregularities of young pulsars with $\tau_{\rm c}<10^5$ years are
dominated by the recovery from previous glitch events (Hobbs et al. 2010). In many cases the NS recovers to the
spin rate prior to the glitch event and thus the glitch event can be removed from the data satisfactorily
without causing significant residuals over model predictions. However in some cases, glitches can cause
permanent changes to both $P$ and $\dot P$ of the NS, which cannot always be removed from the data
satisfactorily, for instance, if the glitch has occurred before the start of the data taking and/or long term
recoveries from glitch events cannot be modeled satisfactorily. These changes can be modeled as a permanent
increase of the surface dipole magnetic field of the NS; consequently some of these NSs may grow their surface
magnetic field strength gradually this way and eventually have surface magnetic field strength comparable to
that of magnetars over a life time of 10$^{5-6}$ years (Lin \& Zhang 2004). These are the first studies that
linked the timing irregularities of pulsars with the long term evolution of magnetic fields of NSs.

The timing irregularities of the second type have time scales of years to decades and thus are normally referred
to as timing noise (Hobbs et al. 2004; 2010). Hobbs et al. (2004, 2010) carried out so far the most extensive
study of the long term timing irregularities of 366 pulsars. Besides ruling out some timing noise models in
terms of observational imperfections, random walks, and planetary companions, Hobbs et al. (2004, 2010)
concluded that the observed $\ddot{\nu}$ values for the majority of pulsars are not caused by magnetic dipole
radiation or by any other systematic loss of rotational energy, but are dominated by the amount of timing noise
present in the residuals and the data span. Some of their main results that are modeled and interpreted in this
work are: (1) All young pulsars have $\ddot{\nu} > 0$; (2) Approximately half of the older pulsars have
$\ddot{\nu} > 0$ and the other half have $\ddot{\nu} < 0$; (3) Quasi-oscillatory timing residuals are observed
for many pulsars; and (4) The value of $\ddot{\nu}$ measured depends upon the data span and the exact choice of
data processed.

Proper understanding of these important results can lead to better understanding of the internal structures and
the magnetospheres of NSs. Since the spin evolution of pulsars cannot be caused by any systematic loss of
rotational energy (Hobbs et al. 2004; 2010), it is natural to introduce some oscillatory parameters. Indeed,
some of the observed periodicities have been suggested as being caused by free-precession of the NS (Stairs et
al. 2000). Some strong quasi-periodicities have been identified recently as due to abrupt changes of the
magnetospheric regulation of NSs, perhaps due to varied particle emissions (Lyne et al. 2010). It was suggested
that such varied magnetospheric particle emissions is also responsible for the observed long term timing noise
(Liu et al. 2011). Alternatively it has been suggested that the propagation of Tkachenko waves in a NS may
modulate its moment of inertia, and thus produce the observed periodic or quasi-periodic timing noise (Tkachenko
1966; Ruderman 1970; Haskell 2011). Because all of the above processes can only cause periodic or quasi-period
modulations to the observed spin of NSs, all these modulations can be modeled phenomenologically as some sort of
oscillations of the observed surface magnetic field strengths of NSs. It is thus reasonable to attribute the
observed wide-spread quasi-oscillatory timing residuals to the oscillations of the magnetic fields of NSs. As
shown in this work, the oscillating magnetic fields can also naturally explain the fact that approximately half
of the older pulsars have $\ddot{\nu} > 0$ and the other half have $\ddot{\nu} < 0$, when the long term magnetic
field decay does not dominate the spin evolutions of pulsars. This is the main advantage of this model, and thus
justifies invoking oscillating magnetic fields for modeling the long term spin evolutions of pulsars.

In this work we attempt to use the known statistical properties of the spin evolution of the pulsars to test
some phenomenological models for the evolution of the dipole magnetic field of a NS. In sections 2-5, we test
the following three models: (1) the standard magnetic dipole radiation model with constant magnetic field, (2)**
long-term exponential decay with an oscillatory component and (3) long-term power-law decay modulated with
oscillations. We find the last model is the only one compatible with the data, with the power-law index
$\alpha>0.5$ and the oscillation amplitude of around $5\times 10^{-5}$ for an oscillation period of several
years. Finally we discuss the physical implication of our results, summarize our results and make conclusions
and discussions, including predictions of the model for further studies. All timing data of pulsars used in this
work are taken from Hobbs et al. (2010).

\section{Testing the standard magnetic dipole radiation model}
\begin{figure}
\center{
\includegraphics[angle=0,scale=0.40]{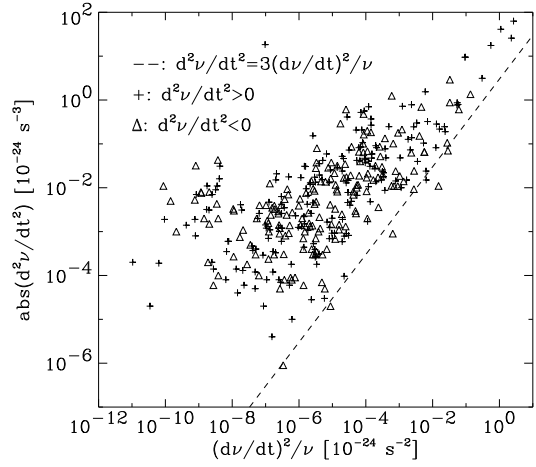}}
\caption{Observed correlation between $\ddot \nu$ and $3\dot{\nu}^2/\nu$. The prediction of the standard
magnetic dipole radiation model, i.e., $\ddot \nu=3\dot{\nu}^2/\nu$ is shown as the dashed line, which
under-predicts significantly the magnitudes of $\ddot \nu$ for most pulsars and also cannot explain $\ddot
\nu<0$ for nearly half of the pulsars.} \label{Fig:1}
\end{figure}

Assuming pure magnetic dipole radiation as the braking mechanism for a pulsar's spin-down, we have,
\begin{equation}\label{dipole}
I\Omega \dot \Omega  =-\frac{(BR^3)^2}{6c^3}{\Omega ^4},
\end{equation}
or,
\begin{equation}\label{nu_dot}
\dot \nu  = -AB^2\nu^3,
\end{equation}
where $A=\frac{(2\pi R^3)^2}{6c^3I}$ is normally assumed to be a constant, $B$ is its dipole magnetic field at
its magnetic pole, $R$ is its radius, $I$ is its moment of inertia. Assuming $\dot B=0$, we have,
\begin{equation}\label{nu_ddot0}
\ddot \nu  = 3\dot{\nu}^2/\nu.
\end{equation}
This model thus predicts $\ddot \nu>0$, against the fact that many pulsars show $\ddot \nu<0$. This is a major
failure of this model, or any other models with monotonic spin-down, as pointed out in Hobbs et al. (2010). In
Fig.~\ref{Fig:1}, we plot the comparison between the observed $\ddot \nu$ and that predicted by
Eq.~({\ref{nu_ddot0}}). It is clear that this model under-predicts the amplitudes of $\ddot \nu$ by several
orders of magnitudes for most pulsars.

\begin{figure}
\center{
\includegraphics[angle=0,scale=0.40]{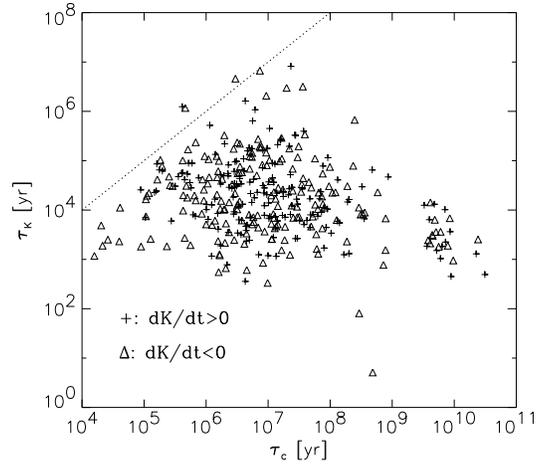}}
\caption{Correlation between the time scale of $K$ variation, $\tau_{K}=K/|{\dot{K}}|$, and the characteristic
age, $\tau_{\rm c}$, of pulsars; $\tau_{K}=\tau_{\rm c}$ is shown as the dashed line. All pulsars show clear
variation of $K$.}\label{Fig:2}
\end{figure}
Assuming $\dot B\ne0$, we have,
\begin{equation}\label{nu_ddot1}
\ddot \nu  = 3\dot{\nu}^2/\nu +2\dot{\nu}\dot{B}/B.
\end{equation}
From the above analysis, it is obvious that the second term in Eq.~({\ref{nu_ddot1}}) must dominate the observed
$\ddot \nu$ (i.e., $3\dot{\nu}^2/\nu \ll |2\dot{\nu}\dot{B}/B|$), for both its magnitude and sign. Define
$K=AB^2$, we have,
\begin{equation}\label{k_dot}
\dot K=\frac{3\dot{\nu}^2-\ddot{\nu}\nu}{\nu^4}.
\end{equation}
We define the evolution time scale of $K$, $\tau_{K}=K/|{\dot{K}}|=B/2|{\dot{B}}|$, we have,
\begin{equation}\label{k_dot}
\tau_{K}=|\frac{\dot{\nu}\nu}{3\dot{\nu}^2-\ddot{\nu}\nu}|.
\end{equation}
Fig.~\ref{Fig:2} shows the observed $\tau_{K}$ as a function of $\tau_{\rm c}$. Clearly $\tau_{K}\ll\tau_{\rm
c}$ for most pulsars. This suggests that if the measured $\ddot{\nu}$ values are caused by magnetic field
evolution, then the evolution must be significant for most pulsars in the sample. Fig.~\ref{Fig:2} also shows
that for all young pulsars with $\tau_{\rm c}< 10^5$~yr, $\dot{K}<0$, i.e., their magnetic moment field decays
with time. For the rest of pulsars, there are almost equal numbers of pulsars with positive or negative
$\dot{K}$. If one assumes that most pulsars are born with some common properties and they then evolve to have
the observed statistical properties, it is then necessary that at least some pulsars should have both
$\dot{K}<0$ and $\dot{K}>0$ at different times, i.e., $K$ oscillates.

We therefore model phenomenologically the evolution of the observed dipole magnetic field of a pulsar by a decay
modulated by the dominated component of oscillations,
\begin{equation}\label{b_decay1}
B(t)= B_{\rm d}(t)(1+k\sin(\phi+2\pi\frac{t}{T})),
\end{equation}
where $t$ is the pulsar's age, $k$, $\phi$ and $T$ are the amplitude, phase and period of the oscillating
magnetic field, respectively. Sometimes it might be necessary to take multiple oscillation components since
multiple peaks are often seen in the power spectra of the timing residuals of many pulsars (Hobbs et al. 2010).
Assuming that all the oscillation time scale is much less than the decay time scale (as will be discussed and
justified later), i.e., $T\ll B_{\rm d}/\dot{B}_{\rm d}$, we have,
\begin{equation}\label{b_decay2}
\dot{B}\simeq \dot{B}_{\rm d}+B_{\rm d} f\cos(\phi+2\pi\frac{t}{T}),
\end{equation}
where $f=2\pi k/T$. For simplicity here we only consider the amplitude of the oscillation term to model the
statistical properties of the timing noise of pulsars, i.e.,
\begin{equation}\label{b_decay3}
\dot{B}= \dot{B}_{\rm d}\pm fB_{\rm d}.
\end{equation}
Since $\dot{B}<0$ can produce $\ddot\nu>0$, a viable model should at least have $|\dot{B}_{\rm d}|/B_{\rm d}>f$
for young pulsars and {\it vice versa} for old pulsars, in order to produce positive $\ddot\nu$ for young
pulsars, but either positive or negative $\ddot\nu$ for old pulsars with almost equal chances, as discussed
above.

\section{Testing Exponential Decay Model with Young Pulsars}

\begin{figure}
\center{
\includegraphics[angle=0,scale=0.40]{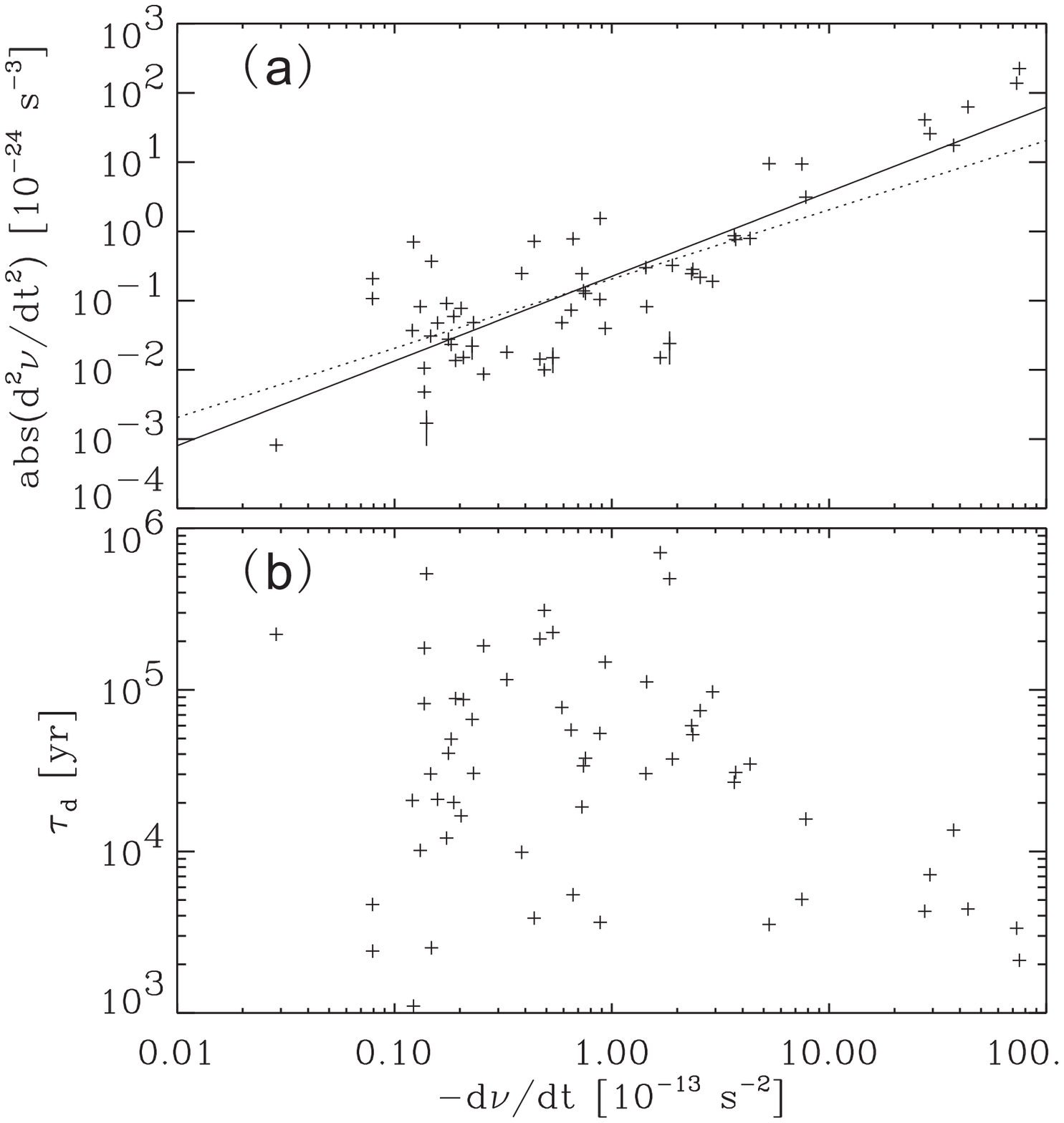}}
\caption{Correlation between $-\dot \nu$ and $\ddot \nu$ (panel (a)) and the exponential decay constant
$\tau_{\rm d}$ (panel (b)) for young pulsars with $\tau_{\rm c}\leq 2\times 10^6$~yr and $\ddot \nu>0$. In panel
(a), the dotted and dashed lines are the best-fit prediction of the exponential decay model, and the best fit
with $\ddot \nu\propto -\dot{\nu}^\gamma$ ($\gamma=1.22$). Panel (b) shows that a single distribution of
$\tau_{\rm d}$ is inconsistent with data.} \label{Fig:3}
\end{figure}

Assuming $B_{\rm d}=B_0 \exp(-t/\tau_{\rm d})$, Eq.~(\ref{b_decay3}) becomes,
\begin{equation}\label{exp}
\dot{B}=B_{\rm d}(-1/\tau_{\rm d}\pm f).
\end{equation}
Substituting Eq.~(\ref{exp}) into Eq.~(\ref{nu_ddot1}) and ignoring its first term (since $3\dot{\nu}^2/\nu \ll
|2\dot{\nu}\dot{B}/B|$), we have,
\begin{equation}\label{ddot_exp}
\ddot{\nu}\simeq -2\dot{\nu}(1/\tau_{\rm d}\pm f).
\end{equation}
Since $\ddot{\nu}>0$ for most young pulsars, we assume $1/\tau_{\rm d}>f$, i.e., $\ddot{\nu}\simeq
-2\dot{\nu}/\tau_{\rm d}$, or, $\tau_{\rm d}\simeq-2\dot{\nu}/\ddot{\nu}$ for young pulsars. In
Fig.~\ref{Fig:3}, we show the observed $\ddot{\nu}$ and the inferred $\tau_{\rm d}$ as a function of the
observed $\dot{\nu}$, for young pulsars with $\tau_{\rm c}\leq10^6$~yr and $\ddot{\nu}>0$. Clearly a single
value or distribution of $\tau_{\rm d}$ cannot describe the data adequately. Another difficulty of this model is
that $\ddot{\nu}<0$ for the nearly half of the pulsars cannot be explained with the same assumption of
$1/\tau_{\rm d}>f$. We thus conclude that a simple exponential decay model is not favored by data.

\section{Testing Power-law Decay Model with Young Pulsars}

\begin{figure}
\center{
\includegraphics[angle=0,scale=0.40]{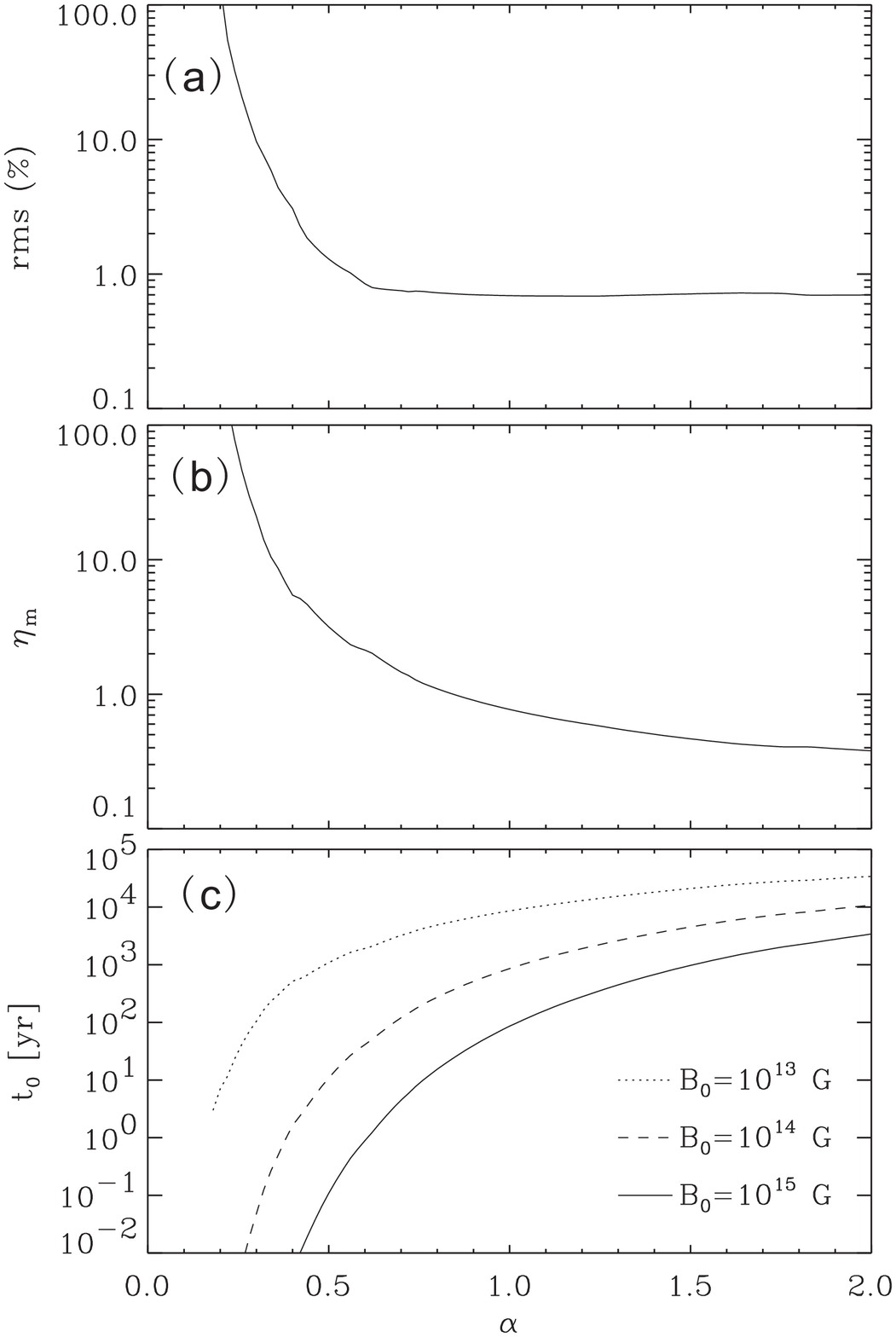}}
\caption{Relations between $\alpha$ and rms of the fit (panel (a)), $\eta_{\rm m}$ (the median value of $\eta$;
(panel (b)), and $t_0$ (the beginning time of power-law decay; panel (c)) for different initial values of $B_0$,
for young pulsars with $\tau_{\rm c}\leq 2\times 10^6$~yr and $\ddot \nu>0$. The data requires $\alpha\geq
0.5$.} \label{Fig:4}
\end{figure}
Assuming $B_{\rm d}=B_0 (t/t_{0})^{-\alpha}$, Eq.~(\ref{b_decay3}) becomes,
\begin{equation}\label{powerlaw}
\dot{B}=B_{\rm d}(-\alpha/t\pm f).
\end{equation}
Substituting Eq.~(\ref{powerlaw}) into Eq.~(\ref{nu_ddot1}) and ignoring its first term (since $3\dot{\nu}^2/\nu
\ll |2\dot{\nu}\dot{B}/B|$), we have,
\begin{equation}\label{ddot_p1}
\ddot{\nu}\simeq -2\dot{\nu}(\alpha/t\pm f),
\end{equation}
where $t$ is the real age of the pulsar. Without other information about $t$, here we replace it with the
magnetic age of a pulsar $t=t_0(B_0/B)^{(1/\alpha)}$; we then have,
\begin{equation}\label{ddot_p2}
\ddot{\nu}\simeq \eta(-\dot{\nu})^{1+\beta}/\nu^{3\beta}\pm 2\dot{\nu}f,
\end{equation}
where $\beta=1/2\alpha$, $\eta=(3.3\times10^{19}/B_0)^{2\beta}/\beta t_0$ and
$B=3.3\times10^{19}\sqrt{P\dot{P}}$~G is assumed.

Since $\ddot{\nu}>0$ for most young pulsars, we assume $\alpha/t>f$, i.e.,
\begin{equation}\label{ddot_p3}
\ddot{\nu}\simeq \eta(-\dot{\nu})^{1+\beta}/\nu^{3\beta},
\end{equation}
for young pulsars. For the $i$-th pulsar, we have,
\begin{equation}\label{eta}
\eta_i=\ddot{\nu}_i/(-\dot{\nu})^{1+\beta}\nu^{-3\beta},
\end{equation}
We then calculate fit the root mean squares (rms) for each value of $\beta=1/2\alpha$,
\begin{equation}\label{rms}
{\rm rms}=\frac{1}{n\eta_{\rm m}}\sqrt{\sum(\eta_i-\eta_{\rm m})^2},
\end{equation}
where $n$ is the number of pulsars and $\eta_{\rm m}$ is the median value of all $\eta_i$. In
Fig.~\ref{Fig:4}(a) we show rms as a function of $\alpha$, indicating that $\alpha\geq 0.5$ is required, but its
upper limit is unconstrained. In Fig.~\ref{Fig:4}(a, b) we also show the ranges of $\eta_{\rm m}$ and the
corresponding $t_0$ and $B_0$.
\begin{figure}
\center{
\includegraphics[angle=0,scale=0.40]{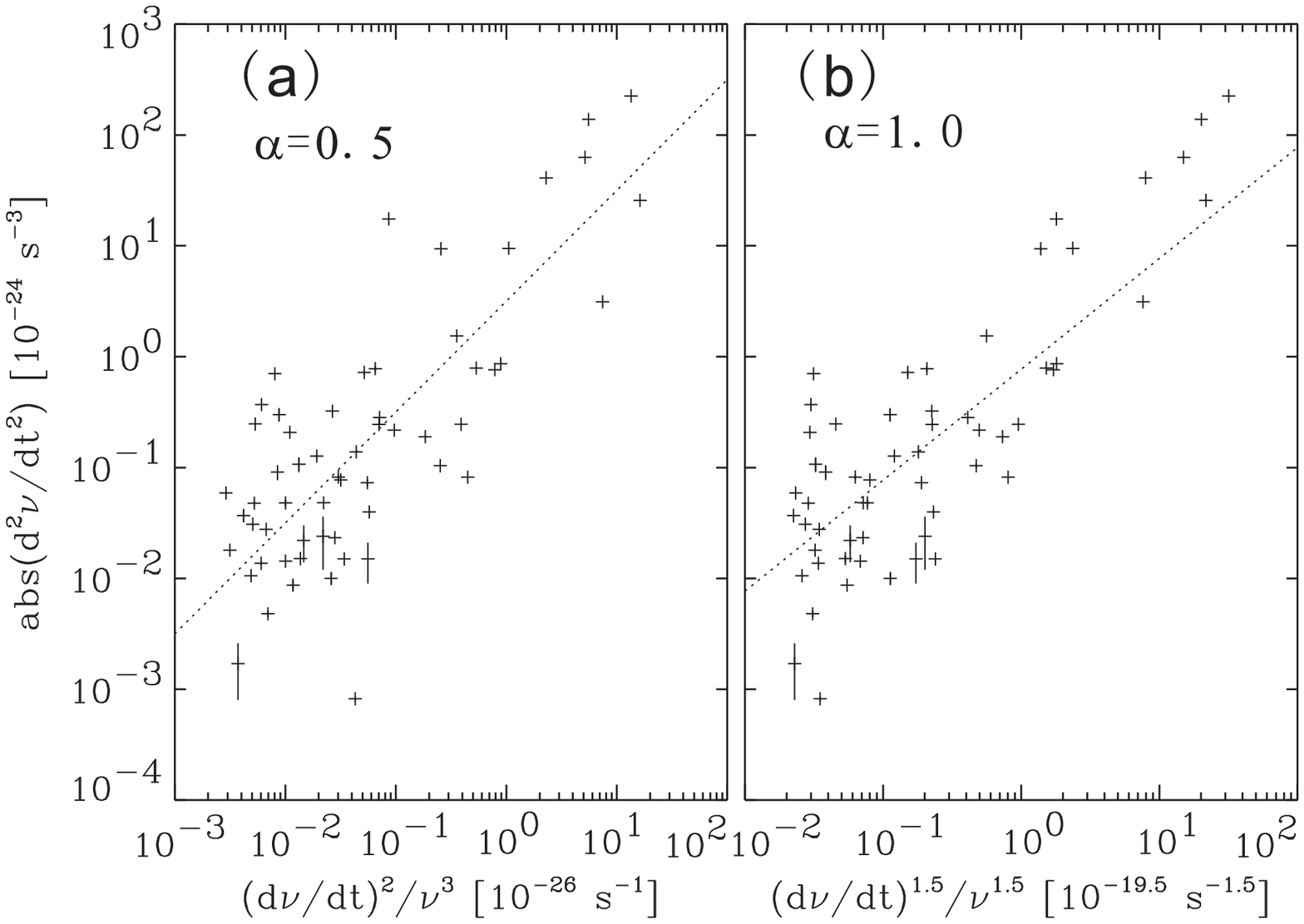}}
\caption{Correlations between $\ddot{\nu}$ and $(-\dot{\nu})^{1+\beta}/\nu^{3\beta}$ with $\alpha=0.5$ (panel
(a)) and $\alpha=1.0$ (panel (b)) for young pulsars with $\tau_{\rm c}\leq 2\times 10^6$~yr and $\ddot \nu>0$.
The dotted lines are the best-fit of $\ddot{\nu}=\eta(-\dot{\nu})^{1+\beta}/\nu^{3\beta}$.} \label{Fig:5}
\end{figure}

The uncertainty in determining the exact value of $\alpha$ is due to the unknown real ages of these pulsars;
using the magnetic ages as their ages introduces a strong coupling between the parameters in the magnetic field
decay model and their ages. This is manifested in the correlations of $\alpha$ with rms, $\eta_{\rm m}$, $t_0$
and $B_0$. Nevertheless, their ranges are reasonable with $\alpha\geq0.5$. In Fig.~\ref{Fig:5}(a, b), we show
the comparison between the predictions of Eq.~(\ref{ddot_p3}) and data with $\alpha=0.5$ and $\alpha=1.0$,
respectively. In both cases the model can describe the data reasonably well. We thus conclude that a simple
power-law decay model is favored by data.

\section{Testing the Model of Power-law Decay Modulated by Oscillations with All Pulsars}

Eq.~(\ref{ddot_p2}) shows that the oscillation term $f$ dominates both the signs and magnitudes of $\ddot{\nu}$
for old pulsars with $t>\alpha/f$; this explains naturally why there are almost equal numbers of pulsars with
$\ddot{\nu}>0$ or $\ddot{\nu}<0$. This agrees with the observation that a single old pulsar can be observed to
show either $\ddot{\nu}>0$ or $\ddot{\nu}<0$, depending on which segment of data is used (Hobbs et al. 2010). We
also confirmed this with Monte-Carlo simulations in a preliminary study (Figs.~5 and 6 in Zhang \& Xie 2012); a
more detailed study will be presented elsewhere (Xie \& Zhang 2012). Fig.~\ref{Fig:4}(a) shows that rms remains
almost constant with $\alpha>1$; we thus restrict our analysis in the following for only two cases: $\alpha=0.5$
and $\alpha=1.0$.

\begin{figure}
\center{
\includegraphics[angle=0,scale=0.40]{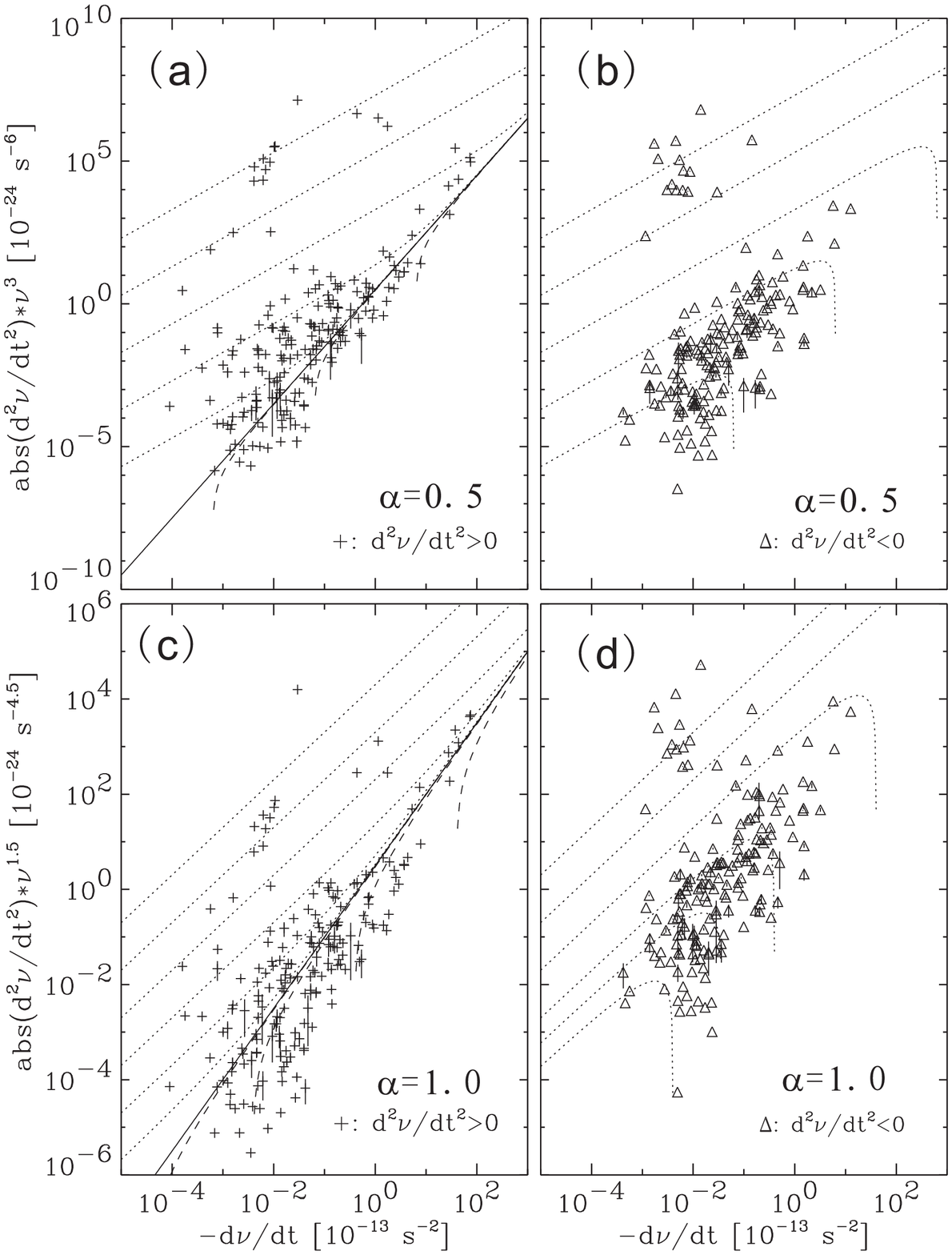}}
\caption{Correlations between $\dot\nu$ and $\ddot{\nu}\nu^{3\beta}$ for all pulsars with $\alpha=0.5$ (panels
(a, b)) and $\alpha=1.0$ (panels (c, d)), in comparison with model predictions with different values of $f$. The
solid lines correspond to $f=0$. The left two panels show data with $\ddot{\nu}>0$, where the dotted and dashed
lines are for $\ddot{\nu}=\eta(-\dot{\nu})^{1+\beta}/\nu^{3\beta}- 2\dot{\nu}f$ (since $\dot \nu <0$) and
$\ddot{\nu}=\eta(-\dot{\nu})^{1+\beta}/\nu^{3\beta}+ 2\dot{\nu}f$, respectively. The right two panels show data
with $\ddot{\nu}<0$ and model predictions of $\ddot{\nu}=\eta(-\dot{\nu})^{1+\beta}/\nu^{3\beta}+ 2\dot{\nu}f$.
The values of $f$ are given in Figs.~(\ref{Fig:7}) and (\ref{Fig:8}).} \label{Fig:6}
\end{figure}

\begin{figure}
\center{
\includegraphics[angle=0,scale=0.40]{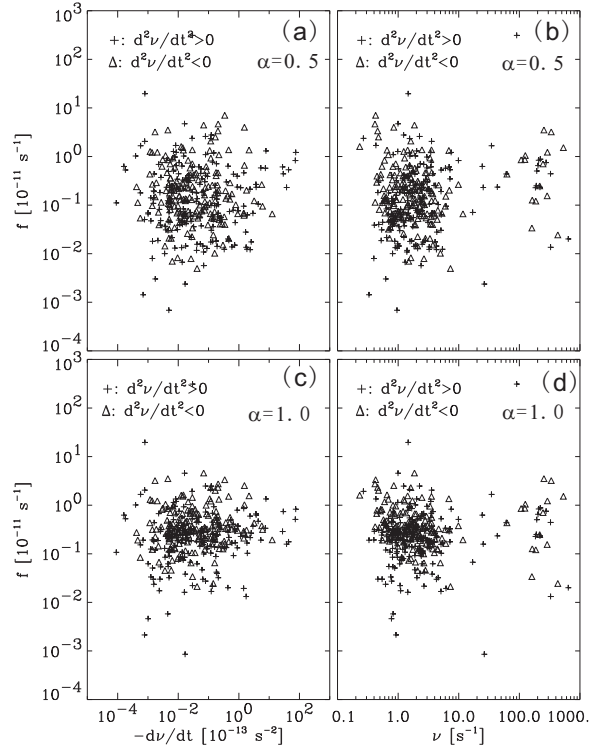}}
\caption{Correlations between $f$ and $-\dot\nu$ (panels (a, c)) and $\nu$ (panels (b, d)), for  $\alpha=0.5$
(panel (a)) and $\alpha=1.0$, respectively. No significant correlation is found.} \label{Fig:7}
\end{figure}

Eq.~(\ref{ddot_p2}) can rewritten as,
\begin{equation}\label{ddot_p4}
\ddot{\nu}\nu^{3\beta}\simeq \eta(-\dot{\nu})^{1+\beta}\pm 2\dot{\nu}\nu^{3\beta}f,
\end{equation}
suggesting that $f\ne 0$ causes deviations to the expected correlation between $\ddot{\nu}\nu^{3\beta}$ and
$\dot{\nu}$, as shown in Fig.~\ref{Fig:6}. Choosing the same range of $f$, the data for both positive and
negative $\ddot{\nu}$ can be explained; $\eta$ is taken as $\eta_{\rm m}$ determined from young pulsars with
$\ddot{\nu}>0$ and $\tau_{\rm c}\leq2\times10^6$~yr, for $\alpha=0.5$ and $\alpha=1.0$, respectively. For each
pulsar in the sample, $f$ can also be determined from Eq.~(\ref{ddot_p2}) or Eq.~(\ref{ddot_p4}), as shown in
Fig.~\ref{Fig:7}; clearly $f$ is not correlated with either $\dot{\nu}$ or ${\nu}$ (we also find that  $f$ is
not correlated with either $\ddot{\nu}$ or $\tau_{\rm c}$, which is not shown here for brevity). This indicates
that the statistical properties of the timing noise of all kinds of pulsars can be explained with the same
oscillation model of similar oscillation amplitudes.

\begin{figure}
\center{
\includegraphics[angle=0,scale=0.40]{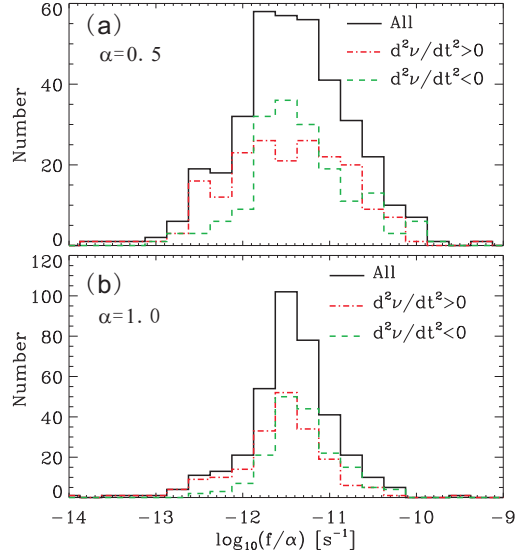}}
\caption{Distributions of $f/\alpha$ of all pulsars (black solid lines), pulsars with positive $\ddot{\nu}$ (red
dot-dashed lines) or negative $\ddot{\nu}$ (green dashed lines), for $\alpha=0.5$ (panel (a)) and $\alpha=1.0$
(panel (b)), respectively. The median values of $f/\alpha$ is about $3\times 10^{-12}$ in both panels.}
\label{Fig:8}
\end{figure}

In Fig.~\ref{Fig:8}, the distributions of $f/\alpha$, for all pulsars with positive or negative $\ddot{\nu}$,
are shown in the cases of $\alpha=0.5$ and $\alpha=1.0$, respectively. All distributions show well-defined peaks
and consistent to each other and the peak value of $f/\alpha$ is around $3\times 10^{-12}$; we also find that
choosing different values of $\alpha$ with $0.5<\alpha<2.0$ always gives similar $f/\alpha$. Finally in
Fig.~\ref{Fig:8} we show the comparisons between the observed $\ddot{\nu}$ and that predicted by
Eq.~(\ref{ddot_p2}) with $f/\alpha=3\times 10^{-12}$ for $\alpha=0.5$ and $\alpha=1.0$, respectively. This
simple model can describe the general properties of the data over about nine orders of magnitudes of
$\ddot{\nu}$.

\begin{figure}
\center{
\includegraphics[angle=0,scale=0.40]{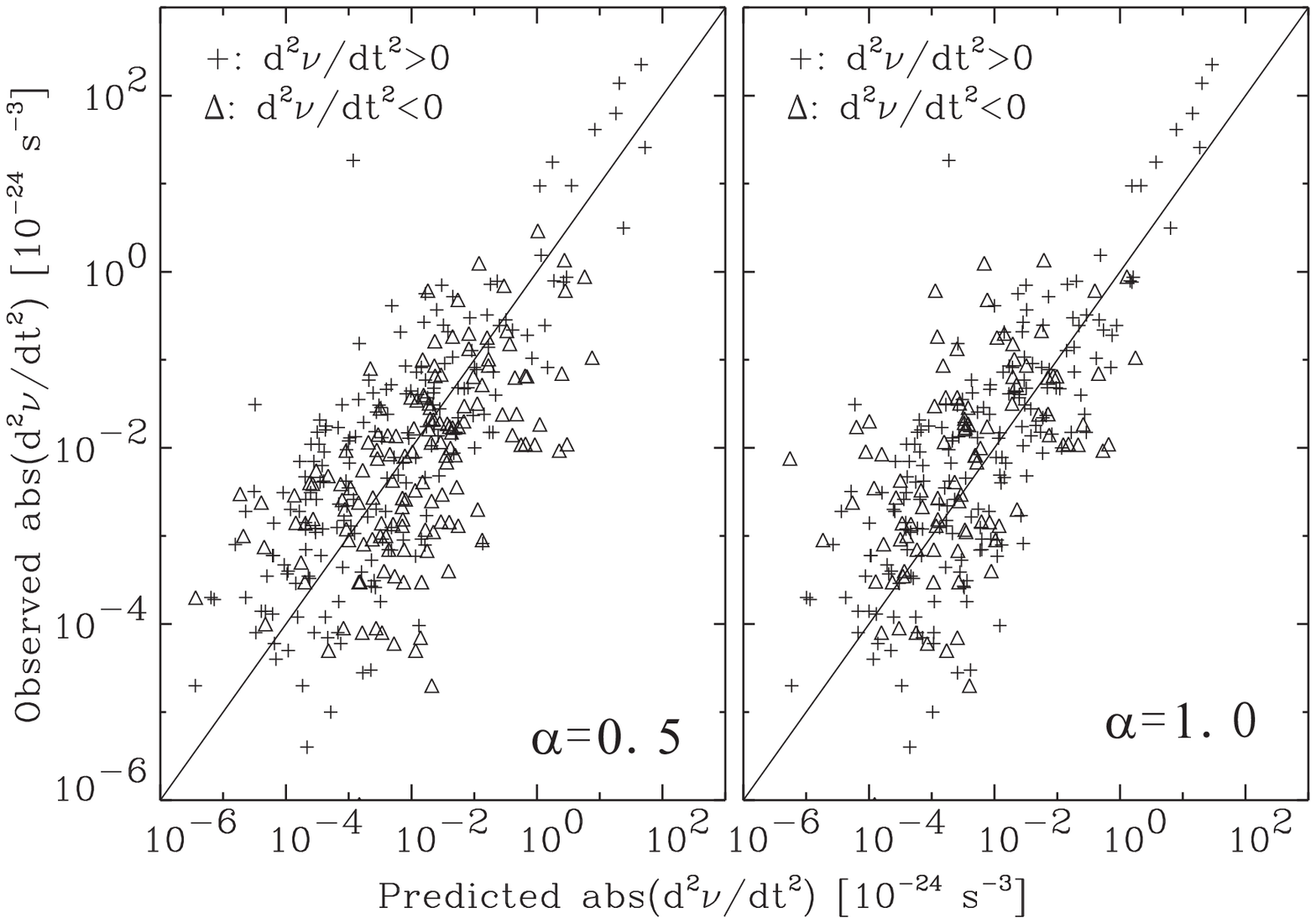}}
\caption{Comparisons between the observed $\ddot{\nu}$ and that predicted by $\ddot{\nu}=
\eta(-\dot{\nu})^{1+\beta}/\nu^{3\beta}\pm 2\dot{\nu}f$ with $f/\alpha=3\times 10^{-12}$ for $\alpha=0.5$ (left
panel) and $\alpha=1.0$ (right panel).} \label{Fig:9}
\end{figure}

\section{Physical Implications}
\subsection{The braking index}
Eq.~(\ref{k_dot}) can be rewritten as,
\begin{equation}\label{k_dot1}
\dot{K}=\frac{\ddot{\nu}}{\nu^3}(\frac{3}{n_{\rm b}}-1),
\end{equation}
where $n_{\rm b}=\ddot{\nu}\nu/\dot{\nu}^2$ is the conventionally defined braking index. Eq.~(\ref{k_dot1})
means that $\dot{K}>0$ when $n_{\rm b}<3$, including $n_{\rm b}<0$ when $\ddot{\nu}<0$. This agrees with many
previous works reporting that $n_{\rm b}<3$ indicates magnetic field {\it increase}, and {\it vice versa} (e.g.,
Blandford \& Romani 1988; Chanmugam \& Sang 1989; Manchester \& Peterson 1989; Lyne et al. 1993; Lyne et al.
1996; Johnston \& Galloway 1999; Wu et al. 2003; Lyne 2004; Chen \& Li 2006; Livingstone et al. 2006;
Middleditch et al. 2006; Livingstone et al. 2007; Yue et al. 2007; Espinoza et al. 2011). However, these works
have reported very few cases that $n_{\rm b}$ is measured reliably.

\begin{figure}
\center{
\includegraphics[angle=0,scale=0.40]{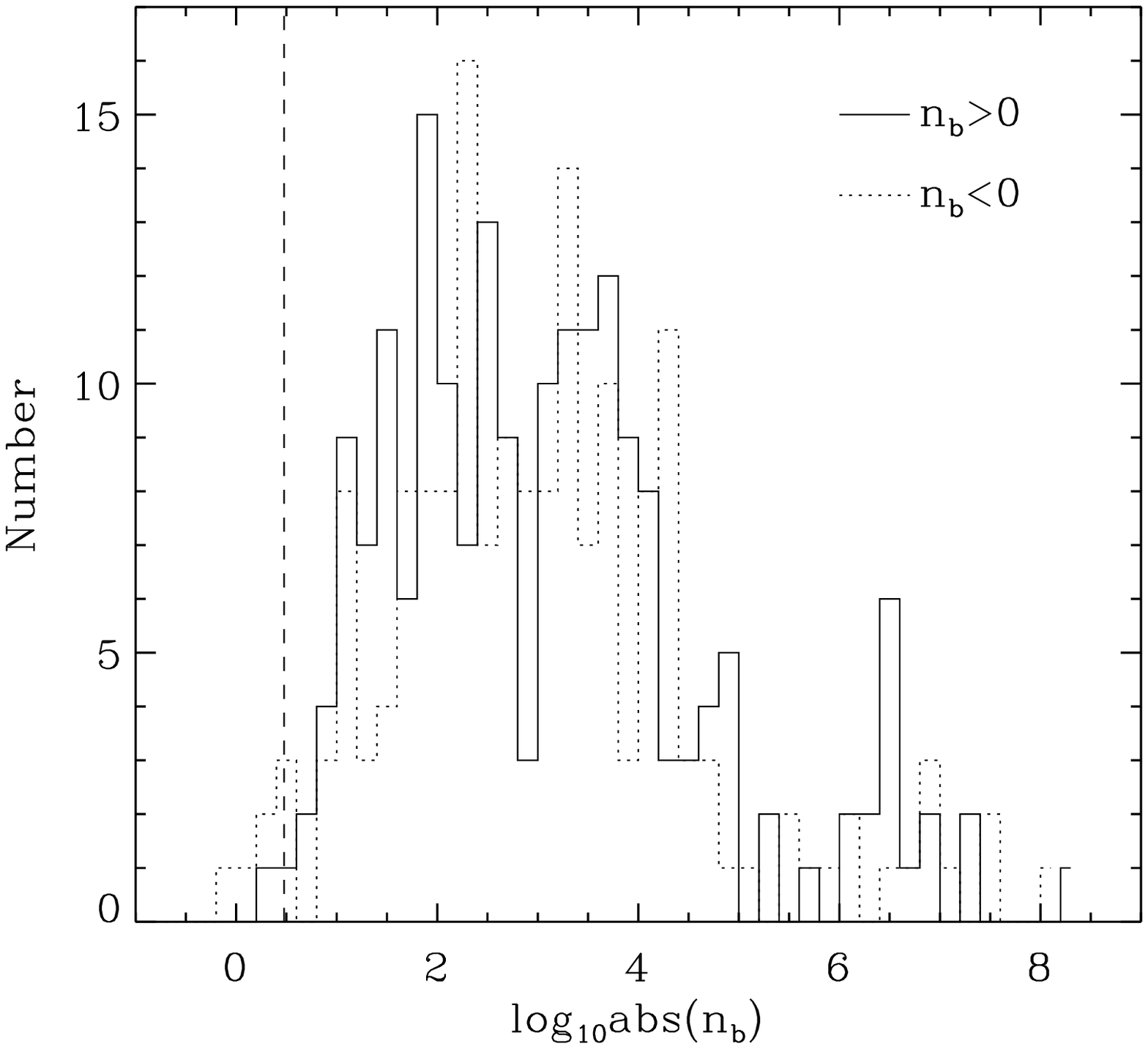}}
\caption{Distributions of the braking index $n_{\rm b}=\ddot{\nu}\nu/\dot{\nu}^2$. The solid and dotted lines
show $n_{\rm b}>0$ and $n_{\rm b}<0$, respectively. The vertical dashed line indicates $n_{\rm b}=3$. The
numbers of positive or negative values of $n_{\rm b}$ are 193 and 173, respectively. The numbers of $n_{\rm
b}>3$ or  $n_{\rm b}<3$ are 192 and 174, respectively.} \label{Fig:10}
\end{figure}

In Fig.~\ref{Fig:10}, we show the distribution of $n_{\rm b}$ in this sample. There is no clustering around
$n_{\rm b}=3$ at all, and the values positive $n_{\rm b}$ and the absolute values of negative $n_{\rm b}$
exhibit almost identical distributions. The numbers of positive or negative values of $n_{\rm b}$ are 193 and
173, respectively. Similarly the numbers of $n_{\rm b}>3$ or $n_{\rm b}<3$ (including all negative $n_{\rm b}$)
are 192 and 174, respectively. Statistically these numbers are indistinguishable. Therefore about half of
pulsars in this sample are observed with increasing or decreasing magnetic fields.

\subsection{Magnetic field decay}
Several physical mechanisms have been proposed for magnetic field decay in NSs, e.g., ohmic dissipation, Hall
effect, and ambipolar diffusion. We can therefore have (Goldreich \& Reisenegger 1992; Heyl \& Kulkarni 1998),
\begin{equation}
\frac{dB}{dt} = -B(\frac{1}{t_{\rm ohmic}}+\frac{1}{t_{\rm ambip}}+\frac{1}{t_{\rm hall}}),
\end{equation}
where $t_{\rm ohmic}\sim2\times10^{11}\frac{L_5^2}{T_8^2}(\frac{\rho}{\rho_{\rm nuc}})^3~{\rm yr}$, $t_{\rm
ambip}^{\rm s}\sim3\times10^{9}\frac{L_5^2T_8^2}{B_{12}^2}~{\rm yr}$ (for the solenoidal component), $t_{\rm
ambip}^{\rm ir}\sim\frac{5\times10^{15}}{T_8^6B_{12}^2}(1+5\times10^{-7}L_5^2T_8^8)~{\rm yr}$ (for the
irrotational component), and $t_{\rm hall}\sim5\times10^{8}\frac{L_5^2T_8^2}{B_{12}}(\frac{\rho}{\rho_{\rm
nuc}})~{\rm yr}$, where $B_{12}$ is the surface magnetic field strength in units of 10$^{12}$~G, $T_8$ denotes
the core temperature in units of $10^8$~K, $\rho_{\rm nuc}\equiv2.8\times10^{14}$~g~cm$^{-3}$ is the nuclear
density, and $L_5$ is a characteristic length scale of the flux loops through the outer core in units of
$10^5$~cm.

Assuming $T_8$ does not vary or only varies slowly with time (Pons et al. 2009), the ohmic dissipation will
produce an exponential decay of the magnetic field, $B= B_{0}e^{-t/\tau_{\rm d}}$. The other two mechanisms all
produce power-law decay, $B= B_{0}(t/t_0)^{-\alpha}$, where $\alpha=1$ or $\alpha=0.5$ for the Hall effect or
ambipolar diffusion (for $t\ll10^{12}$~yr), respectively. We recently found that this model with constant core
temperatures of NSs can explain well the observed age mismatch of NSs discussed above, and the ambipolar
diffusion effect seems to be dominant, though the statistics is rather limited (Zhang \& Xie 2011; Xie \& Zhang
2011)

As shown in Fig.~(\ref{Fig:3}) and discussed in Sect. 3, the simple exponential decay cannot describe the data
for young pulsars with $\tau_{\rm c}\leq 10^6$~yr. Therefore a constant core temperature ($T_8=const$) ohmic
dissipation model for all magnetic field decay is ruled out. However, for a cooling NS, i.e., $T_8$ decreases
with time, the ohmic dissipation model would result in an increasing $\tau_{\rm d}$ with time, in rough
agreement with that shown in Fig.~(\ref{Fig:3})(b). Indeed, an increasing $\tau_{\rm d}$ with time would also
make the first term in Eq.~(\ref{ddot_exp}) less important or negligible for old pulsars, thus explaining why
there are almost equal number of pulsars with either positive or negative $\ddot{\nu}$. Of course an increasing
$\tau_{\rm d}$ with time may be able to mimic a power-law decay mathematically.

As shown in Figs.~(\ref{Fig:4}-\ref{Fig:9}) and discussed in Sects. 4 and 5, a power-law decay with $\alpha\geq
0.5$ can explain all the observed statistical properties of the timing noise of pulsars. Unfortunately the exact
value or even the upper limit of $\alpha$ cannot be constrained by the data, though $\alpha\approx 1$ is
preferred. For both the Hall effect and the ambipolar diffusion (for $t\ll10^{12}$~yr), a cooling NS implies
faster magnetic field decay, and thus increasing $\alpha$ with time.

We therefore conclude that the current data do not conflict the above discussed model of magnetic field decay of
NSs, but also cannot constrain precisely how these effects regulate their magnetic field decay.

\subsection{Magnetic field oscillations}

As demonstrated in Sect. 5, invoking an oscillation component with $f/\alpha\sim3\times10^{-12}$~s$^{-1}$, both
the amplitude and signs of $\ddot{\nu}$ can be well reproduced for all pulsars. Quasi-periodic oscillations with
time scales of the order of years in the timing residuals of pulsars are found to be quite common (Hobbs et al.
2010). As we have shown recently, such quasi-periodic oscillations can be re-produced by an oscillating dipole
magnetic field of a NS with an oscillation period of several years (Zhang \& Xie 2012). Taking $T=3$~years and
$\alpha=1$, we have the oscillation amplitude of the observed magnetic field $k\sim5\times 10^{-5}$.

Assuming that these oscillations are caused by precessing of the NS and the angle between its magnetic and spin
axes is 60 degrees, $k\sim5\times 10^{-5}$ requires a precession angle of about 0.003 degree, which is much
smaller than the inferred 0.3 degree precession angle for PSR~B1828$-$11 from its periodic timing properties
(Stairs et al. 2000). Therefore PSR~B1828$-$11 is probably an extreme example showing observed oscillating
magnetic fields, though our result suggests that the same phenomenon exists in essentially all pulsars. Of
course the apparent oscillations of their observed magnetic fields may also be caused by periodic or
quasi-periodic changes of their magnetospheres (Lyne et al. 2010; Liu et al. 2010), or even their moments of
inertia caused by, for example, the Tkachenko waves in NSs. Regardless which mechanism is responsible for the
oscillations, the mechanism appear to work in the same way for all types of NSs, since the distribution of
$f/\alpha$ is quite narrow (especially when $\alpha=1$), and $f$ does not appear to be correlated with any
observed parameters of pulsars.
\section{Summary, conclusion and discussion}
In this work we have tested models of magnetic field evolution of NSs with the statistical properties of their
timing noise reported by Hobbs et al. (2010). In all models the magnetic dipole radiation is assumed to dominate
the spin-down of pulsars; therefore different models of their magnetic field evolution lead to different
properties of their spin-down. Our main results in this work are summarized as follows:

(1) We tested the standard model of pulsar's magnetosphere with constant magnetic field. We found that this
model under-predicts the second derivatives of their spin frequencies ($\ddot{\nu}$) by several orders of
magnitudes for most pulsars (Fig.~(\ref{Fig:1})). This model also fails to predict $\ddot{\nu}<0$, which is
observed in nearly half of the old pulsars. The standard model of pulsar's magnetosphere with constant magnetic
field is thus ruled out, and $\ddot{\nu}$ is found to be dominated by the varying magnetic field.

(2) Statistically about half of the pulsars in this sample show increasing or decreasing magnetic fields. The
magnetic field strength of most pulsars are found to either increase or decrease with time scales significantly
shorter than their characteristic ages (Fig.~(\ref{Fig:2})).

(3) We constructed a phenomenological model of the evolution of the magnetic fields of NSs, which is made of a
long term decay modulated by short term oscillations.

(4) A simple exponential decay is not favored by the apparent correlation of the inferred decay constant with
their characteristic ages for young pulsars (Fig.~(\ref{Fig:3})(b)) and fails to explain the fact that
$\ddot{\nu}$ is negative for roughly half of old pulsars.

(5) A simple power-law decay, however, can explain all the observed statistical properties of $\ddot{\nu}$
(Figs.~(\ref{Fig:5} and \ref{Fig:6})) and infer a narrow distribution of an oscillation amplitude of about
$5\times 10^{-5}$ with a preferred decay index of about unity (Figs.~(\ref{Fig:7} and \ref{Fig:8})).

We thus conclude that long term magnetic field and short term oscillations are ubiquitous in all NSs and that
the data support the model consisting of a power-law decay modulated by short term oscillations for the
evolution of their magnetic fields. However we can only constrain the decay index to be $\alpha\geq0.5$
(Figs.~(\ref{Fig:4}) and (\ref{Fig:8})(b)).

Our results are consistent with the models of magnetic field decay we discussed, in which ohmic dissipation,
Hall effect, and ambipolar diffusion are possible effects causing magnetic field decay. However, the uncertainty
in the decay index $\alpha$ makes it difficult to constrain further or distinguish between these effects. The
problem in determining $\alpha$ is the unavailability of the true ages of pulsars. With the true age $t$
available for some pulsars, $\alpha$ can be determined directly with Eq.~(\ref{ddot_p1}); $f=0$ can be safely
assumed for young pulsars. Unfortunately only several pulsars in the sample of Hobbs et al (2010) have the ages
of their supernova remnants available, thus preventing a statistically meaningful determination of $\alpha$ this
way. We thus suggest that reliable determination of the true ages of a large sample of young NSs is needed, in
order to constrain further the physical mechanisms of the magnetic field decay of NSs. This can also provide an
independent test of our model.

The main prediction of this model is that the spin evolution of each pulsar should follow,
\begin{equation}\label{ddot_p2}
\ddot{\nu}(t)\simeq 2\dot{\nu}(t)(-\frac{\alpha}{t}+\sum f_i\cos(\phi_i+2\pi\frac{t}{T_i})),
\end{equation}
where both $\ddot{\nu}(t)$ and $\dot{\nu}(t)$ should be time-resolved, rather than the averaged value
conventionally reported over the whole data span. Hobbs et al. (2010) have shown that both the value and sign of
the averaged $\ddot{\nu}$ depend on the time span used, which can be reproduced with our model presented here
(Zhang \& Yi 2012). However a clear-cut and robust test of our model is to compare the prediction of the above
equation. As discussed above, the first or the second term in the above equation may be ignored for old
millisecond pulsars or young pulsars, respectively. The current on-going programs of intensive long term precise
timing monitoring of pulsars should produce the required data for validating our model. In doing so, both the
decay index and the oscillating properties of each individual pulsar can be determined, which in turn can
provide important insights into the physics of NSs.

 \acknowledgments
We appreciate discussions with and comments from E. P. J. van den Heuvel, S. Tsuruta, J. L. Han, W. W. Tian, X.
P. Zheng, R. X. Xu, F. J. Lu, H. Tong, D. Chen, X. F. Wu, C. M. Zhang, and J. F. Liu. The anonymous referee is
thanked for his/her comments and suggestions. SNZ acknowledges partial funding support by the National Natural
Science Foundation of China under grant Nos. 11133002, 10821061, 10725313, and by 973 Program of China under
grant 2009CB824800.


\begin{thebibliography}{}
\bibitem[Aguilera et al.(2008)]{2008ApJ...673L.167A} Aguilera, D.~N., Pons, J.~A., \& Miralles, J.~A.\ 2008, \apjl, 673, L167
\bibitem[Bhattacharya et al.(1992)]{1992A&A...254..198B} Bhattacharya, D., Wijers, R.~A.~M.~J., Hartman, J.~W., \& Verbunt, F.\ 1992, \aap, 254, 198
\bibitem[Blandford \& Romani(1988)]{1988MNRAS.234P..57B} Blandford, R.~D., \& Romani, R.~W.\ 1988, \mnras, 234, 57P
\bibitem[Chanmugam \& Sang(1989)]{1989MNRAS.241..295C} Chanmugam, G., \& Sang, Y.\ 1989, \mnras, 241, 295
\bibitem[Chen \& Li(2006)]{2006A&A...450L...1C} Chen, W.~C., \& Li, X.~D.\ 2006, \aap, 450, L1
\bibitem[Espinoza et al.(2011)]{2011ApJ...741L..13E} Espinoza, C.~M., Lyne, A.~G., Kramer, M., Manchester, R.~N., \& Kaspi, V.~M.\ 2011, \apjl, 741, L13
\bibitem[Geppert et al.(1999)]{1999A&A...345..847G} Geppert, U., Page, D., \& Zannias, T.\ 1999, \aap, 345, 847
\bibitem[Goldreich\& Reisenegger(1992)]{1992ApJ...395..250G} Goldreich, P., \& Reisenegger, A.\ 1992, \apj, 395, 250
\bibitem[Gonthier et al.(2002)]{2002ApJ...565..482G} Gonthier, P.~L., Ouellette, M.~S., Berrier, J., O'Brien, S., \& Harding, A.~K.\ 2002, \apj, 565, 482
\bibitem[Gunn\& Ostriker(1970)]{1970ApJ...160..979G} Gunn, J.~E., \& Ostriker, J.~P.\ 1970, \apj, 160, 979
\bibitem[Guseinov et al.(2004)]{2004IJMPD..13.1805G} Guseinov, O.~H., Ankay, A., \& Tagieva, S.~O.\ 2004, International Journal of Modern Physics D, 13, 1805
\bibitem[Han(1997)]{1997A&A...318..485H} Han, J.~L.\ 1997, \aap, 318, 485
\bibitem[Harding\& Lai(2006)]{2006RPPh...69.2631H} Harding, A.~K., \& Lai, D.\ 2006, Reports on Progress in Physics, 69, 2631
\bibitem[Haskell(2011)]{2011PhRvD..83d3006H} Haskell, B.\ 2011, \prd, 83, 043006
\bibitem[Heyl\& Kulkarni(1998)]{1998ApJ...506L..61H} Heyl, J.~S., \& Kulkarni, S.~R.\ 1998, \apjl, 506, L61
\bibitem[Hobbs et al.(2004)]{2004MNRAS.353.1311H} Hobbs, G., Lyne, A.~G.,Kramer, M., Martin, C.~E., \& Jordan, C.\ 2004, \mnras, 353, 1311
\bibitem[Hobbs et al.(2010)]{2010MNRAS.402.1027H} Hobbs, G., Lyne, A.~G., \& Kramer, M.\ 2010, \mnras, 402, 1027
\bibitem[Holt\& Ramaty(1970)]{1970Natur.228..351H} Holt, S.~S., \& Ramaty, R.\ 1970, \nat, 228, 351
\bibitem[Johnston \& Galloway(1999)]{1999MNRAS.306L..50J} Johnston, S., \& Galloway, D.\ 1999, \mnras, 306, L50
\bibitem[Lin\& Zhang(2004)]{2004ApJ...615L.133L} Lin, J.~R., \& Zhang, S.~N.\ 2004, \apjl, 615, L133
\bibitem[Liu et al.(2011)]{2011ChPhL..28a9701L} Liu, X.-W., Na, X.-S., Xu, R.-X., \& Qiao, G.-J.\ 2011, Chinese Physics Letters, 28, 019701
\bibitem[Livingstone et al.(2006)]{2006ApJ...647.1286L} Livingstone, M.~A., Kaspi, V.~M., Gotthelf, E.~V., \& Kuiper, L.\ 2006, \apj, 647, 1286
\bibitem[Livingstone et al.(2007)]{Livingstone}Livingstone, M. A., Kaspi, V. M., Gavriil, F. P., et al. 2007, Ap\&SS, 308, 317
\bibitem[Lorimer(2011)]{2011heep.conf...21L} Lorimer, D.~R.\ 2011, High-Energy Emission from Pulsars and their Systems, 21
\bibitem[Lyne et al.(1975)]{1975MNRAS.171..579L} Lyne, A.~G., Ritchings, R.~T., \& Smith, F.~G.\ 1975, \mnras, 171, 579
\bibitem[Lyne et al.(1993)]{1993MNRAS.265.1003L} Lyne, A.~G., Pritchard, R.~S., \& Graham-Smith, F.\ 1993, \mnras, 265, 1003
\bibitem[Lyne et al.(1996)]{1996Natur.381..497L} Lyne, A.~G., Pritchard, R.~S., Graham-Smith, F., \& Camilo, F.\ 1996, \nat, 381, 497
\bibitem[Lyne(2004)]{2004IAUS..218..257L} Lyne, A.~G.\ 2004, Young Neutron Stars and Their Environments, IAU Symposium no. 218,  (Eds., Camilo, F. \& Bryan M. Gaensler, B. M.), 218, 257
\bibitem[Lyne et al.(2010)]{2010Sci...329..408L} Lyne, A., Hobbs, G., Kramer, M., Stairs, I., \& Stappers, B.\ 2010, Science, 329, 408
\bibitem[Manchester \& Peterson(1989)]{1989ApJ...342L..23M} Manchester, R.~N., \& Peterson, B.~A.\ 1989, \apjl, 342, L23
\bibitem[Middleditch et al.(2006)]{2006ApJ...652.1531M} Middleditch, J., Marshall, F.~E., Wang, Q.~D., Gotthelf, E.~V., \& Zhang, W.\ 2006, \apj, 652, 1531
\bibitem[Ostriker\& Gunn(1969)]{1969ApJ...157.1395O} Ostriker, J.~P., \& Gunn, J.~E.\ 1969, \apj, 157, 1395
\bibitem[Pacini(1969)]{1969Natur.224..160P} Pacini, F.\ 1969, \nat, 224, 160
\bibitem[Pons et al.(2009)]{2009A&A...496..207P} Pons, J.~A., Miralles, J.~A., \& Geppert, U.\ 2009, \aap, 496, 207
\bibitem[Popov et al.(2010)]{2010MNRAS.401.2675P} Popov, S.~B., Pons, J.~A., Miralles, J.~A., Boldin, P.~A., \& Posselt, B.\ 2010, \mnras, 401, 2675
\bibitem[Regimbau \& de Freitas Pacheco(2001)]{2001A&A...374..182R} Regimbau, T., \& de Freitas Pacheco, J.~A.\ 2001, \aap, 374, 182
\bibitem[Ridley\& Lorimer(2010)]{2010MNRAS.404.1081R} Ridley, J.~P., \& Lorimer, D.~R.\ 2010, \mnras, 404, 1081
\bibitem[Ruderman(1970)]{r70} Ruderman, M. 1970, \nat, 225, 619
\bibitem[Ruderman(2005)]{2005esns.conf...47R} Ruderman, M.\ 2005, NATO ASIB Proc.~210: The Electromagnetic Spectrum of Neutron Stars, 47
\bibitem[Tauris \& Konar(2001)]{2001A&A...376..543T} Tauris, T.~M., \& Konar, S.\ 2001, \aap, 376, 543
\bibitem[Tkachenko(1966)]{1966JETP...23.1049T} Tkachenko, V.~K.\ 1966, Soviet Journal of Experimental and
Theoretical Physics, 23, 1049
\bibitem[Wu et al.(2003)]{2003A&A...409..641W} Wu, F., Xu, R.~X., \& Gil, J.\ 2003, \aap, 409, 641
\bibitem[Xie \& Zhang(2011)]{2011ASPC..451..253X} Xie, Y., \& Zhang, S.\ 2011, Astronomical Society of the Pacific Conference
Series, 451, 253 (arXiv:1110.3869)
\bibitem[Xie\& Zhang(2012)]{} Xie, Y., \& Zhang, S.-N.\ 2012, in preparation
\bibitem[Yue et al.(2007)]{2007AdSpR..40.1491Y} Yue, Y.~L., Xu, R.~X., \& Zhu, W.~W.\ 2007, Advances in Space Research, 40, 1491
\bibitem[Zhang \& Xie(2011)]{2011ASPC..451..231Z} Zhang, S., \& Xie, Y.\ 2011, Astronomical Society of the Pacific Conference Series, 451, 231 (arXiv:1110.3154)
\bibitem[Zhang \& Xie(2012)]{2012arXiv1202.1123Z} Zhang, S.-N., \& Xie, Y.\ 2012, to appear in IJMPD (arXiv:1202.1123)
\end{thebibliography}
\end{document}